# Generative Inverse Estimation of 3D Atomic Coordination from Near-Edge Spectra via Equivariant Diffusion Models

[1]Ren Okubo, [1,2]Yu Fujikata, [1]Izumi Takahara, and [1]Teruyasu Mizoguchi

[1]Institute of Industrial Science, The University of Tokyo,
Komaba, Meguro, Tokyo 153-8505, Japan

[2]Science & Innovation Center, Mitsubishi Chemical Corporation,
Kanagawa, 227-8502, Japan

**Abstract**
Extracting 3D atomic coordinates from spectroscopic data is a longstanding inverse problem. We present an equivariant diffusion model that generates site-specific 3D structures directly from near-edge spectra (ELNES/XANES). Trained on Si-O crystals, the model achieves radial accuracy comparable to Extended X-ray Absorption Fine Structure (EXAFS) (RMSD ~0.06 Å) but with superior coordination number precision (errors < 4.3% vs. EXAFS ~20%). Crucially, it reconstructs full 3D geometries including bond angles, overcoming the limitations of 1D radial distribution analysis. The model demonstrates robust out-of-distribution generalization, accurately predicting local structures in amorphous systems despite being trained exclusively on crystalline lattices. Application to experimental O $K$-edge spectra from $\alpha$-quartz validates practical applicability. This generative approach outperforms template matching and establishes automated, quantitative 3D structure determination from spectroscopic data.

**Introduction**
The creation of novel materials underpins innovation across electronics, aerospace, automotive engineering, energy technologies, and healthcare[1–3]. At the heart of this progress lies precise knowledge of local three-dimensional (3D) atomic coordination: bond lengths, coordination numbers, and bond angles govern macroscopic properties in decisive ways. Achieving atomic-level control over these structural features has therefore become a central goal in materials development, driving demand for analytical methodologies capable of resolving 3D coordination environments with atomic-scale precision.

Among existing techniques, scanning transmission electron microscopy (STEM) represents one of the most effective approaches for atomic-resolution structural analysis. In STEM, a focused electron probe is scanned across a specimen to form images by detecting transmitted or scattered electrons. While atomic-resolution images can be obtained by converging the probe to dimensions smaller than atomic column spacing, the resulting data are limited to two-dimensional projections. Although TEM/STEM tomography can reconstruct three-dimensional information by acquiring images at multiple tilt angles, beam-induced damage, limited accessible tilt ranges, and reconstruction artifacts often compromise the completeness and accuracy of structural analysis, particularly for complex materials[4,5].

Complementary to imaging, spectroscopic techniques provide access to local coordination information. Extended energy-loss fine structure (EXELFS) in electron microscopy and its X-ray counterpart, extended X-ray absorption fine structure (EXAFS), have long served as cornerstone methods for local structure determination[6–10]. These techniques analyze oscillations in spectra arising from scattering of excited electrons or X-rays by the surrounding atomic environment. Fourier transformation yields radial distribution functions (RDFs) from which interatomic distances and coordination numbers can be extracted. While EXAFS achieves impressive bond distance "fitting" precision (typically 0.01–0.02 Å for known structural models), it suffers from well-documented limitations: coordination number determination is notoriously unreliable with typical uncertainties of ~20% due to strong correlations between coordination number and Debye-Waller factors[11,12], and RDFs inherently lack angular information about bond orientations. Moreover, EXELFS and EXAFS require data acquisition over an extended high-energy range beyond the absorption edge, which is experimentally challenging and time-consuming, particularly in electron microscopy where beam damage and specimen drift become critical concerns.

In contrast to extended fine structure, the near-edge region of core-level excitation spectra—electron energy-loss near-edge structure (ELNES) in electron microscopy and X-ray absorption near-edge structure (XANES) in X-ray spectroscopy—can be acquired over a narrower energy range, making measurements more experimentally feasible and efficient. These near-edge spectra originate from excitation of core electrons at specific atomic sites and are highly sensitive to local coordination geometry, bond angles, and elemental composition[7,13]. The spectral features encode rich information about the three-dimensional arrangement of atoms surrounding the excited site. However, extracting explicit three-dimensional atomic coordinates from ELNES/XANES remains a fundamental challenge: interpretation has traditionally relied on qualitative comparison with reference spectra or first-principles calculations, limiting throughput and requiring substantial expertise.

Recognizing these challenges, researchers have increasingly turned to data-driven approaches to extract structural information from ELNES and XANES[14–19]. Machine learning methods can learn complex patterns from large datasets to predict properties efficiently. For example, Tanaporn et al.[20] applied random-forest models to XANES spectra to predict local properties such as oxidation states, coordination numbers, and average nearest-neighbor bond lengths. Takahara et al.[21] demonstrated prediction of ground-state electronic structures from core-loss spectra. Kiyohara et al.[22,23] developed neural network models to predict coordination numbers and radial distribution functions from ELNES/XANES, demonstrating that averaged structural descriptors can be inferred from near-edge spectra. While these advances have been significant, existing methods have largely been limited to predicting averaged local-structure descriptors and have not yet achieved generation of full three-dimensional atomic coordinates at specific excitation sites. Recent advances in generative modeling offer new possibilities for structure prediction. Diffusion models, in particular, have achieved groundbreaking results across diverse domains including image generation, computer vision, and natural language processing, and their application to atomic structure generation is advancing rapidly[24–27]. Kwon et al.[28] recently demonstrated conditional generation of amorphous carbon structures using a diffusion model with XANES spectra as conditioning inputs. Their approach successfully predicted descriptor distributions such as RDFs and bond angles that matched target structures. However, their work focused on bulk amorphous carbon in large supercells and evaluated structures using global descriptors, without assessing site-specific three-dimensional correspondence between predicted and target atomic coordinates. Thus, while this pioneering work demonstrated the potential of generative models for spectroscopy-guided structure prediction, the question of whether local three-dimensional atomic coordinates around individual excitation sites can be accurately predicted from ELNES/XANES spectra has remained open.

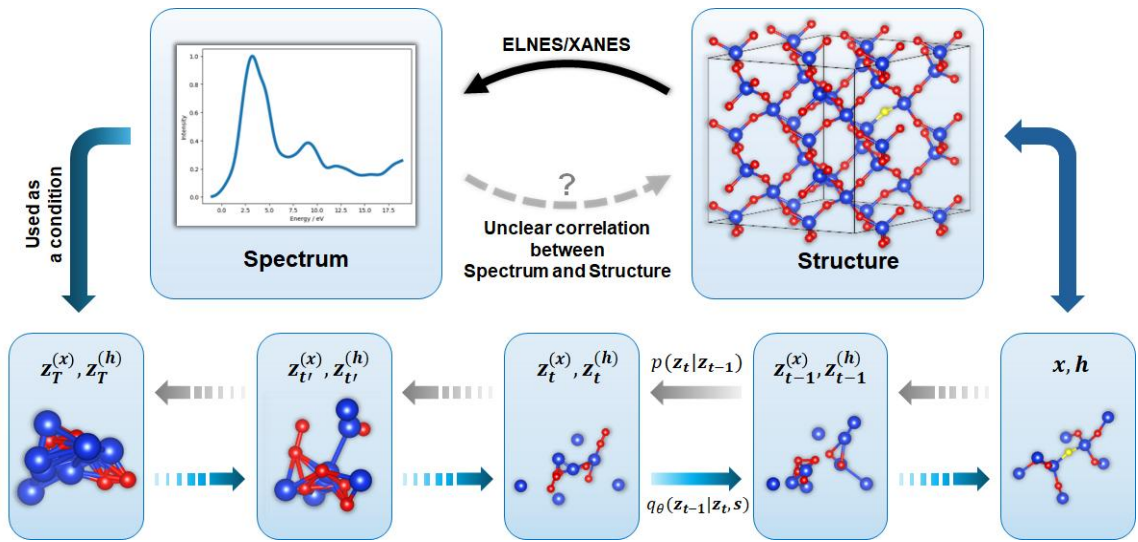

**Figure 1 Overview of spectral-to-structure prediction approach.** While ELNES/XANES spectra encode rich information about local atomic coordination environments, extracting explicit three-dimensional atomic positions has remained a long-standing challenge. Traditional methods like EXAFS provide radial distribution information with bond distance fitting precision of 0.01–0.02Å but suffer from poor coordination number accuracy (~20% uncertainty) and lack direct access to angular information. Our generative diffusion model, conditioned on experimental or calculated ELNES/XANES spectra, directly generates complete three-dimensional atomic coordinates around the absorbing atom. The denoising process begins with Gaussian noise $z_T$ and progressively refines atomic cooridinates, $x$, and node features $h$—comprising atomic species $a$—, and through timesteps ($z_t^{(x)}$, $z_t^{(h)}$, $z_{t'}^{(x)}$, $z_{t'}^{(h)}$, $z_{t-1}^{(x)}$, $z_{t-1}^{(h)}$...) using an equivariant graph neural network, ultimately producing physically realistic 3D atom coordinates consistent with the input spectral condition.

To address this gap, we have developed a conditional diffusion model that generates three-dimensional atomic coordinates in the local coordination environment around individual excited atomic directly from ELNES/XANES spectra (Fig.1). Our approach utilizes a diffusion model with an equivariant graph neural network architecture, explicitly preserving the rotational and translational symmetries inherent to atomic structures. Unlike previous work that predicted global structural descriptors, our method generates explicit three-dimensional positions of atoms surrounding the excitation site, enabling direct comparison with ground-truth structures and quantitative evaluation of coordinate-level prediction accuracy.

We focus on binary Si-O compounds as a model system for several compelling reasons. First, Si-O materials are of significant industrial and technological importance: silica and silicon oxides are fundamental in semiconductor manufacturing, lithium-ion battery anodes, catalysis, and optical devices[29–31]. Second, the O $K$-edge in Si-O compounds exhibit rich spectral features that are highly sensitive to local coordination environments. Third, the system benefits from numerous well-characterized crystal structures with diverse coordination motifs, providing an ideal dataset for training and validation. These characteristics make the Si-O system an optimal platform for demonstrating and rigorously evaluating site-specific three-dimensional coordination generation from ELNES/XANES spectra.

## Results

We trained a diffusion model on 2,067 computed O $K$-edge spectra from 274 Si-O crystal structures, enabling direct generation of three-dimensional atomic coordinates conditioned on spectroscopic input. The training dataset encompassed a compositional range of SiO$_x$ (x=0.5-4.0), though dominated by SiO$_2$ structures (1,842 of 2,067 spectra, Supplementary Fig. 1), reflecting the natural prevalence of silica polytypes in crystalline databases. Computational details for the O $K$-edge spectra are provided in Methods.

To rigorously evaluate our method's predictive power and transferability, we designed a three-tier validation strategy. First, we assessed quantitative accuracy on held-out crystalline test structures, comparing performance against conventional template-matching approaches. Second, we challenged the model with amorphous $SiO_x$ systems (x =0.25-2.0, prepared separately as described in Methods) to test whether a model trained exclusively on crystalline data could reconstruct the local coordination environments in disordered amorphous materials. Finally, we applied the crystalline-trained model to experimental O $K$-edge spectra from $α$-quartz to demonstrate real-world applicability. Together, these benchmarks establish that site-specific three-dimensional atomic coordination can be predicted directly from near-edge spectroscopic data with accuracy competitive to or exceeding traditional structural characterization methods.

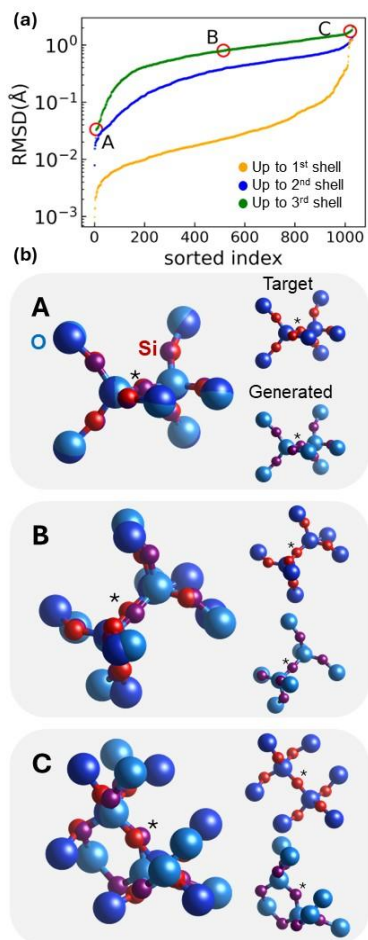

**Figure 2 Root-mean-square distance (RMSD) for Results of Conditional Generation.** (a) RMSD between target and generated structure of all the test data for 1st, 2nd, and 3rd shells. RMSD distribution across 206 test structures. (b) Structures corresponding to the plot A, B, C in (a). The target structure is shown in the upper right of each frame and the generated structure in the lower right; an overlay of these two structures is shown to the left. The asterisk-marked site corresponds to the excited atomic site. Blue and red spheres correspond to Si and oxygen, respectively.

### *Generation of 3D coordination from test spectra*
We first evaluated the model's fundamental capability: can it accurately generate three-dimensional atomic coordinates around excitation sites when conditioned on ELNES/XANES spectra? For each test spectrum, we generated structures by sampling initial coordinates $X_{initial}$, and atomic species $A_{initial}$ from a standard normal distribution, conditioning on the spectrum $S_{condition}$, and running the reverse diffusion process. As prediction targets, we considered coordination environments extending to third-nearest neighbors from the excited oxygen site. Figure 2 presents quantitative validation of the model's structural prediction accuracy across all test spectra. Following alignment of the predicted and target structures using the procedure described in the Methods section, RMSD was calculated separately for atoms within the first, second, and third coordination shells; the results are shown in Figure 2(a). The overall mean RMSD across all atoms within the first three coordination shells is 0.804 Å (80 pm). More revealing, however, is the shell-specific analysis: the first coordination shell achieves a mean RMSD of 0.063 Å (6.3 pm), directly comparable to the typical bond-distance fitting precision of EXAFS analysis (0.01~0.02 Å). This is remarkable considering that EXAFS achieves such precision through iterative refinement against pre-defined structural models, whereas our approach generates atomic coordinates de novo from spectral data alone.

Beyond radial accuracy, the model demonstrates exceptional performance in coordination number prediction, a well-known limitation of conventional EXAFS analysis, which typically suffers from ~20% uncertainties due to correlation effects between coordination number and Debye-Waller factors[11,12]. Our model achieves coordination number prediction errors of only 1.86% and 4.27% for the first and second shell, representing more than one order of magnitude improvement over EXAFS capabilities.

Most critically, unlike any radial distribution technique, our diffusion model simultaneously predicts full three-dimensional atomic arrangements including angle—information that EXAFS/EXELFS cannot provide directly. We note that for structures with nine or more atoms that locally break symmetry, the alignment procedure (Methods) does not guarantee the globally optimal rotation matrix, and slightly smaller RMSDs may be achievable with alternative alignments. Nonetheless, the achieved accuracy across multiple metrics—comparable radial precision to EXAFS (0.063 Å vs 0.01–0.02 Å), order-of-magnitude superior coordination number accuracy (1.86% vs ~20%), and unique 3D coordinate prediction capability—positions spectral-to-structure generation as a powerful complement to, and in some aspects an improvement over, traditional local structure determination methods.

Figure 2(b) shows coordination environments corresponding to three representative cases marked by red circles in Fig. 2(a). For the high accuracy case A, the RMSD of the predicted structure relative to the target was 0.033 Å, indicating that not only distances but also all structural parameters including bond angles agree with the target with high accuracy. Visual inspection of the aligned structures confirms that the central O used for conditioning, the first-shell Si, the second-shell O, and even the third-shell Si almost completely overlap between the target and predicted structures.

For the intermediate-accuracy case B (RMSD 0.797 Å), predictions are highly accurate up to second-nearest neighbors, with Si–O bond lengths well reproduced, whereas bond angles from the second- to third-nearest neighbors deviate from those of the target; nonetheless, even the third-shell oxygens are predicted at relatively close coordinates. In the low accuracy case C (RMSD 1.726 Å), the predicted structure contains Si-Si bonds and six-membered ring motifs absent in the target. Because such motifs occur in a subset of training environments, we infer that spectral conditioning was insufficient in this case, leading to partial generation of features from other spectra. Distance-dependent analysis revealed a general trend: atoms closer to the excited oxygen site are predicted more accurately than those farther away.

Next, we quantitatively evaluated prediction accuracy of elemental species. Figure 3 shows the oxygen fraction in generated versus target structures. We defined accuracy as the proportion of test cases where the oxygen fraction matches exactly. Despite only two species (Si and O) being present, achieving an accuracy of 0.949 indicates very high performance. We also examined relationships among prediction accuracy, compound composition, and training sample frequency. When generated elemental species did not match targets, coordinate RMSDs also tended to be large. Regarding compositional dependence, coordination environments from $SiO_2$—the most abundant composition in the dataset—showed high success rates for species generation, whereas less frequent compositions such as $Si_2O_5$ and $Si_3O_8$ exhibited lower accuracy.

the database (the training split of our Si-O crystal dataset), then assigned the structure corresponding to the minimum MSE. To enable fair comparison, we quantified structural similarity using the Smooth Overlap of Atomic Positions (SOAP) descriptor[32]. We embedded both template-matched and model-generated structures into SOAP vectors and computed their cosine similarity to the target structure. During template matching, we excluded pairings between different sites within the same material to avoid trivial matches.

Result of the comparison is shown in Fig. 4. The horizontal axis is the MSE between the matching spectrum and the target spectrum, while the vertical axis is the difference between similarities: (generative model similarity) minus (template matching similarity). Points in green indicate cases where template matching achieves higher accuracy, whereas points in purple indicate superior performance by our generative model. Remarkably, our model generated structures with higher accuracy than template matching in 52.1% of test cases. The distribution of similarity differences (right panel) peaks sharply near zero with mean 0.001 and standard deviation 0.006. Combined with the 52.1% success rate, these results demonstrate that our generative approach not only matches but frequently surpasses conventional template matching accuracy, while retrieving pre-existing structures.

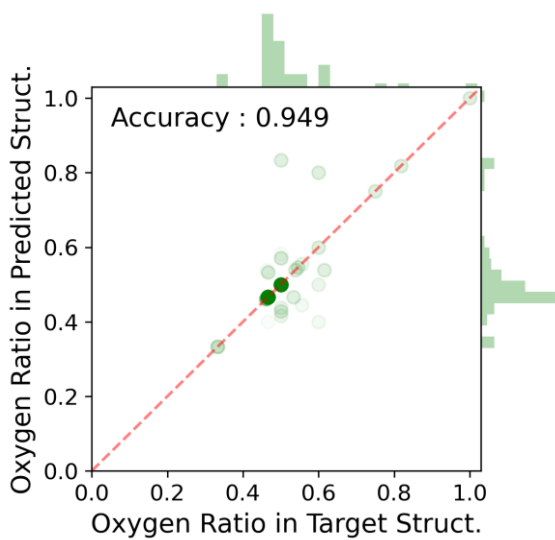

**Figure 3 Prediction accuracy of elemental species.** The accuracy is defined as the proportion of test cases where the oxygen fraction matches exactly.

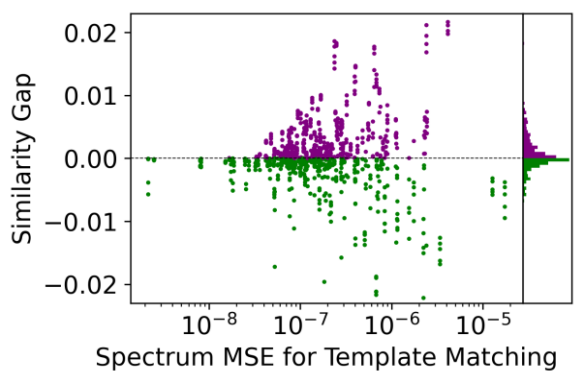

**Figure 4 Comparison with conventional template matching.** The horizontal axis shows the mean squared error (MSE) between the best-matching spectrum from the template database and the target spectrum, where larger values indicate more difficult cases for template matching. The vertical axis shows the structural similarity gap, defined as the difference in SOAP-based cosine similarity to the target structure: [similarity of generative model prediction] minus [similarity of template matching result]. Purple points indicate cases where our generative model achieves higher structural similarity to the target; green points indicate cases where template matching performs better. The right panel shows the distribution of similarity gaps. Our model outperforms template matching in 52.1% of test cases.

*Comparison with conventional template matching*

Having demonstrated accurate coordinate generation, we next asked whether this approach offers advantages over established analysis methods. Template matching—comparing a target spectrum against a database of reference spectra and assigning the structure of the closest match—has been the conventional approach for extracting structural information from ELNES/XANES. We directly compared our generative model against template matching to quantify performance differences. In template matching, we computed the mean squared error (MSE) between each test spectrum and all reference spectra in

*Application to amorphous materials*

A critical question for practical applicability is whether a model trained exclusively on crystalline structures can be generalized to amorphous materials. This question has particular technological relevance: amorphous Si-O materials are used across diverse fields including glass, catalysis, electronics, biomedicine, and surface coating[29,31,33]. Moreover, amorphous $SiO_x$ is regarded as a promising anode material for lithium-ion batteries due to high capacity and silicon resource abundance, with oxygen concentration strongly correlated to battery performance[30]. Fundamental understanding of local coordination in amorphous $SiO_x$ could therefore significantly impact materials design across these applications. To investigate this, we constructed amorphous $SiO_x$ structures (x = 0.25, 0.5, 0.75, 1.0, 1.25, 1.5, 1.75, 2.0) via classical molecular dynamics simulations using the Tersoff potential[34,35]. Preparing the cell consisting of around 200 atoms, we performed cooling from molten states and subsequently applied structural relaxation to obtain amorphous configurations[36]. For each oxygen site in the resulting structures, we computed the core-level excitation spectra under the same conditions described below and used these spectra as conditioning inputs during generation as described in Methods. We did not include any amorphous spectra in the training dataset and ensured that the model was trained solely on crystalline Si-O compounds.

Figures 5(a) and 5(b) show, for amorphous $SiO_{2.0}$ and $SiO_{0.25}$, the RMSD of the three-dimensional coordinates in conditional generation as well as the target and predicted structures. The results for all other compositions x=0.5, 0.75, 1.0, 1.25, 1.5, and 1.75 are summarized in Supplementary Figure 3. It is important to note that the model was trained exclusively on crystalline structures; the amorphous configurations therefore represent a rigorous out-of-distribution generalization test. For $SiO_{2.0}$ and $SiO_{0.25}$, the mean first-shell RMSDs were 0.137 and 0.778Å, respectively.

For amorphous $SiO_{2.0}$, predictions reproduced atoms near the excited oxygen site with relatively high accuracy, and in many cases successfully captured local motifs such as oxygen coordinated by two Si atoms. The elemental species prediction accuracy for amorphous $SiO_{2.0}$ was 84.4% (Supplementary Fig. 1). For amorphous $SiO_{0.25}$, the RMSD distribution exhibited bimodality: one group below 0.3 Å and another around 1 Å. The low-RMSD group corresponds to coordination environments with few atoms near the excited oxygen, a consequence of using a 2.0 Å cutoff radius and including up to third-nearest neighbors. In $SiO_{0.25}$, Si–Si bonds are sometimes not counted as neighboring atoms, leading to smaller coordination environments. While RMSD values were similar across compositions, visual inspection revealed that predictions for $SiO_{2.0}$ typically captured overall structural shapes, whereas those for $SiO_{0.25}$ often failed to reproduce characteristic local motifs such as Si–Si bonds and oxygen coordinated by three Si atoms. The elemental species prediction accuracy for $SiO_{0.25}$ was notably lower at 16.4%, with a tendency to overestimate oxygen content.

Despite having no exposure to disordered structures during training, the model successfully generated reasonable three-dimensional atomic coordinates for amorphous $SiO_{2.0}$ as shown in Figure 5(b), demonstrating a promising degree of transferability. The larger RMSD observed for $SiO_{0.25}$ reflects the more heterogeneous local coordination environments at this composition, which deviate more substantially from the crystalline training distribution. Together, these results suggest that the learned spectral-to-structure mapping captures the features of local coordination geometry that generalize beyond the crystalline domain.

*Understanding performance differences across compositions*
To understand these compositional differences, we visualized training and amorphous spectra via UMAP[37] (Fig. 6 (a)). Amorphous spectra show composition-dependent distributions, shifting continuously as oxygen content varies. Training spectra cluster on the left, often near $SiO_{2.0}$ and $SiO_{1.75}$ amorphous spectra. Indeed, approximately 90% of training data came from $SiO_2$ materials, with the remaining ~10% from compositions such as SiO, $Si_2O_3$, and $SiO_4$ (Supplementary Fig. 2). This explains why the model performs well on $SiO_2$ amorphous spectra, they lie close to the training distribution, while struggling with $SiO_{0.25}$, which lies far from any training examples.

Figure 6(b) and 6(c) compare our model against template matching for amorphous $SiO_{2.0}$ and $SiO_{0.25}$ amorphous spectra, respectively. For $SiO_{2.0}$, both methods perform well due to the similarity of those spectra to the training data, with our model achieving highest similarity in over half of test cases.
For SiO$_{0.25}$, where spectra lie far from the training distribution, our model nevertheless achieved higher similarity in over 70% of test cases. This suggests the model learned robust correlations between spectral features and structures, enabling it to outperform template matching even for challenging cases where the reference database lacks similar spectra. These results demonstrate that the learned spectral-structural relationships can generalize beyond the training distribution, though we anticipate further improvement through training on more diverse compositions.

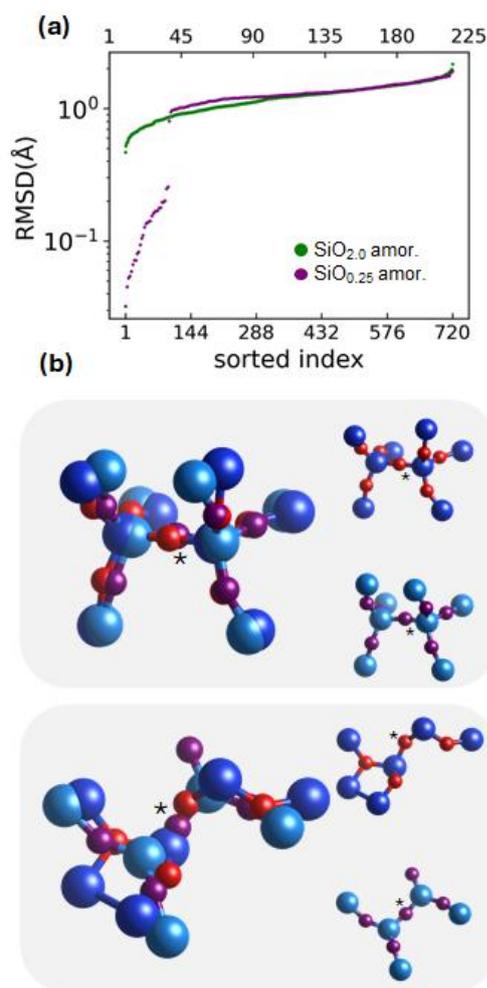

**Figure 5 Conditional Generation for Amorphous (amor).** (a) Coordination evaluation for $SiO_{2.0}$ and $SiO_{0.25}$ amor. Upper x-axis: $SiO_{0.25}$ amor. Lower x-axis: $SiO_{2.0}$ amor. (b) Target structures and predicted structures from spectrum of $SiO_{2.0}$ and $SiO_{0.25}$ amor. The target structure is shown in the upper right of each frame and the generated structure in the lower right; an overlay of these two structures is shown to the left. The asterisk-marked site corresponds to the excited atomic site. Blue and red spheres correspond to Si and oxygen, respectively.

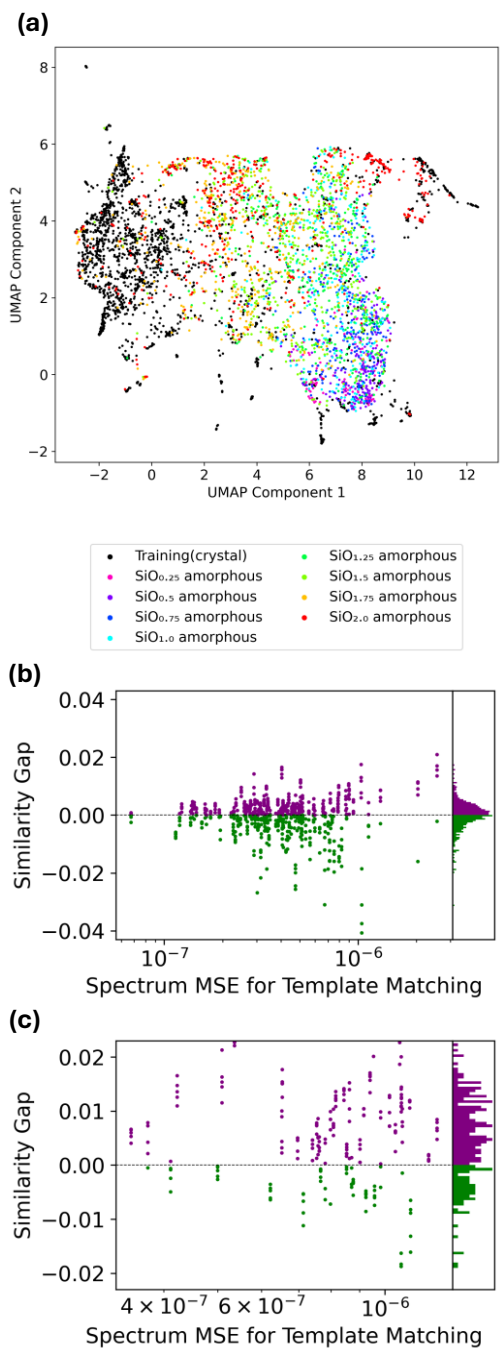

**Figure 6 Relationship between model prediction and data distribution.** (a) Spectrum data distribution created by UMAP. (b) Comparison between template matching and model prediction for SiO$_{2.0}$ amorphous. (c) Comparison between template matching and model prediction for SiO$_{0.25}$ amorphous.

*Applicability for experimental spectrum*
Finally, we examined the applicability of the proposed model to experimental spectra. Here, we consider conditional generation using the experimental O $K$-edge spectrum of $\alpha$-quartz. $\alpha$-quartz was selected in this study as it provides a well-characterized reference system[22,38,39]. In the $\alpha$-quartz crystal, the oxygen sites are crystallographically equivalent, so a single O $K$-edge can be used for coordination-environment prediction.
Figures 7(a)–7(c) show the experimental and calculated O $K$-edge spectrum of $\alpha$-quartz, the radial distribution function centered on the excited oxygen, and the target versus generated structures. The RMSD between the generated and target coordinates was 0.705 Å. In the RDF, the peak positions within approximately 3 Å of the excited oxygen largely coincided, whereas the peak around 4 Å, corresponding to the third-nearest neighbors, was not reproduced with full fidelity. This is consistent with previous work using neural networks to predict RDFs from XANES/ELNES[39], which also reported deviations in the 3–5 Å range. That study identified the near-edge rise region as the primary contributor to predicting average bond lengths and angles. In our case, although the experimental and calculated spectra differ somewhat in the post-edge region (likely due to experimental noise), the near-edge rise regions align well (Fig.7(a)). This suggests that, as in previous studies, the near-edge rise played a major role in our model's distance and angle predictions. We anticipate improved accuracy through finer energy resolution in calculations and smoothing of experimental spectra.
This successful application to experimental data validates that the model can generate meaningful coordination environments from real measurements, not just calculated spectra, establishing practical applicability for materials characterization.

**Discussion**
This work establishes that diffusion models can extract site-specific three-dimensional atomic coordinates directly from ELNES/XANES spectra, a capability that opens new avenues for materials characterization.
The quantitative accuracy achieved by our model merits careful evaluation in the context of established structure determination techniques. For first-shell bond distances, our mean RMSD of 0.063 Å is only 3–6 times larger than the typical "fitting" precision of EXAFS analysis (0.01–0.02 Å), despite addressing a fundamentally different—and arguably more challenging—inverse problem. EXAFS achieves its impressive precision through iterative refinement against pre-defined structural models, whereas our approach generates atomic coordinates de novo from spectral data alone, without any structural priors or candidate models. More significantly, our model demonstrates exceptional performance in coordination number prediction, achieving errors of only 1.86% (first shell) and 4.27% (second shell). This represents more than one order of magnitude improvement over EXAFS analysis, where coordination number uncertainties of ~20% are widely recognized as a fundamental limitation arising from strong correlations between coordination number and Debye-Waller factors. This dramatic improvement transforms coordination number determination from a well-known "pitfall" of EXAFS analysis into a reliable quantitative measurement[12].

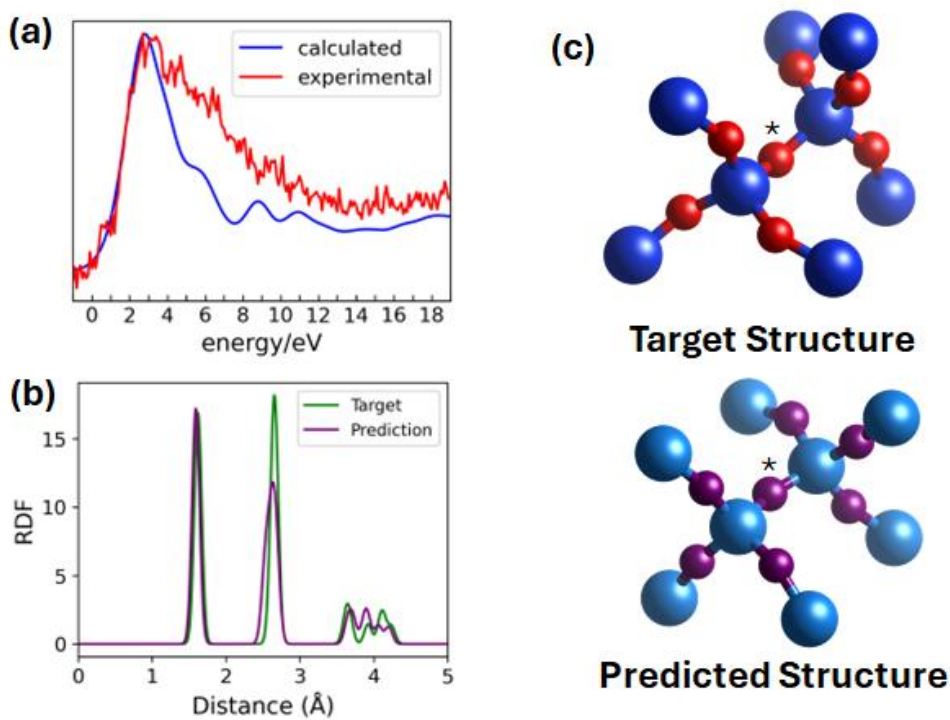

**Figure 7 Conditional generation from experimental spectrum.** (a) Experimental Spectrum of alpha quartz. (b) Radial distribution function of target structure and predicted structure. (c)Target Structure and predicted Structure. The asterisk-marked site corresponds to the excited atomic site. Blue and red spheres correspond to Si and oxygen, respectively.

Most critically, our method uniquely provides complete three-dimensional atomic positions including bond angles and dihedral angles—structural information that EXAFS cannot directly access and EXELFS obtains only through complex multiple-scattering analysis requiring significant structural assumptions. This combination of capabilities—comparable radial accuracy, superior coordination number precision, and unique 3D coordinate prediction—positions our approach not merely as an alternative to EXAFS/EXELFS, but as a complementary tool that addresses key limitations of traditional methods while maintaining competitive performance in their areas of strength.

The method is immediately applicable to several important classes of materials: (1) systems with crystallographically equivalent sites, exemplified here by a-quartz, where all oxygen atoms share identical coordination environments; (2) two-dimensional materials and layered compounds where equivalent sites are arrayed along specific crystallographic directions[40]; (3) grain boundaries and interfaces exhibiting local symmetry; and (4) materials amenable to emerging depth-resolved spectroscopy techniques. For these systems, spectrum-to-structure prediction provides a powerful analytical tool that complements conventional characterization methods.

While acquiring truly site-resolved spectra from individual atoms distributed in three dimensions remains experimentally challenging, the measurement landscape is rapidly evolving. Notable progress has been made using state-of-the-art STEM-EELS, and emerging techniques such as depth-resolved ELNES using high-convergence electron beams are beginning to provide quasi-site-specific information along the depth direction[41]. Beyond electron microscopy, synchrotron-based XANES measurements offer complementary capabilities: while typically providing ensemble-averaged spectra, recent advances in focused X-ray beams and scanning techniques enable spatial resolution approaching tens of nanometers. Moreover, even spatially averaged spectra from materials with local disorder can provide valuable structural information when combined with our generative modeling framework, as demonstrated by the amorphous structure results.

Beyond immediate applications, several clear pathways exist for enhancing model performance and broadening scope. Expanding training datasets to encompass diverse compositions, particularly off-stoichiometric and minority phases, would improve predictions across the full compositional range. Incorporating multi-edge information—for example, combining O $K$-edge with Si $L$-edge and/or $K$-edge, could provide complementary constraints to enhance both coordinate accuracy and elemental species identification. Extension to ternary and higher-order systems would enable analysis of the broader range of technologically relevant materials. Integration of uncertainty quantification would allow the model to indicate prediction confidence, particularly valuable when extrapolating beyond training distributions.

The demonstrated ability to outperform conventional template matching in over half of crystalline test cases and over 70% of amorphous cases, combined with quantitative accuracy approaching EXAFS precision for bond distances and exceeding EXAFS performance for coordination numbers, while generating explicit three-dimensional coordinates rather than retrieving pre-existing structures, represents a fundamental advance. Unlike database lookup methods, our generative approach can produce novel coordination environments, potentially identifying unexpected local structures, defects, or intermediate phases. The successful generalization from crystalline to amorphous materials, despite training exclusively on ordered structures, demonstrates that the model has learned transferable spectral-structural relationships rather than merely memorizing the training set.

Looking forward, integration with high-throughput experimental workflows could enable rapid screening of local coordination in combinatorial materials libraries. Combined with active learning strategies, where the model identifies high-uncertainty predictions to guide targeted measurements, this could create closed-loop characterization systems that accelerate materials discovery. The approach is also directly applicable to operando studies, where tracking local coordination changes during battery cycling, catalytic reactions, or phase transformations could provide unprecedented insight into dynamic processes at the atomic scale.

**Methods**

*Dataset preparation, spectrum calculations and graph representation*

This study targets 274 Si–O compounds registered in the Materials Project[42,43]. Si–O compounds are widely used in glass, batteries, and the semiconductor industry, and their three-dimensional (3D) local coordination environments are closely related to optical properties, ion conductivity, and dielectric characteristics; thus, structural analysis is of industrial importance[29–31,33,44]. For 2,067 symmetrically unique oxygen sites, we constructed a dataset that pairs the ELNES/XANES spectrum at each site with the local coordination structure around that site, and we computed the O $K$-edge. The O $K$-edge lies near 530 eV and can be measured using electron microscopy or soft X-rays. Because the O $K$-edge mainly reflects information on O-$p$ orbitals in the conduction band and on O-Si hybridization in addition to oxygen itself, it is well suited as the model spectrum used in this work.

Electronic-structure calculations for the O $K$-edge were performed using CASTEP[45] based on the plane-wave pseudopotential method, and the oxygen $K$-edge spectra were computed with OptaDOS[46]. For structural relaxation, we employed the GGA–PBE exchange–correlation functional. The plane-wave cutoff energy was uniformly set to 500 eV, and k-points were chosen using a Monkhorst–Pack grid[47] such that the k-point density was 0.03 Å$^{-1}$. Spin polarization was not considered. By generating the pseudopotential for the excited oxygen "on the fly," we calculated the excited-state electronic structure while accounting for the core-hole effect. To avoid interactions between core holes, we used supercells. In spectral calculations, with the Fermi energy set to 0, the number of unoccupied bands was adjusted so that states were included up to 25 eV above the Fermi level in addition to the occupied states. Each spectrum was spline-interpolated to 200 points (0.1 eV steps, -1~19 eV) and normalized to unit area.

We treated, as the coordination environment corresponding to each spectrum, the atomic species and coordinates of atoms up to the third-nearest neighbors from the excited oxygen site, and set these as the targets for generation by the model. Neighboring atoms were defined as sites within a cutoff radius (= 2.0 Å) from each atomic site. Data in which no atoms existed in the first-nearest shell—i.e., environments defined by only a single atom—were excluded from the dataset.

We also generated amorphous structural data for SiO$_x$ (x = 0.25, 0.5, 0.75, …, 2.0) via molecular dynamics (MD) using the Tersoff potential[34–36]. After structural relaxation, ELNES/XANES spectra were computed for these structures under the same conditions as above, and the resulting data were used as test cases to assess the model's applicability to amorphous systems. In the armophous simulation, totally 45, 75, 90, 100, 125, 144, 140, and 144 O $K$-edges from oxygen sites were calculated for x=0.25, 0.5, 0.75, 1.0, 1.25, 1.5, 1.75, 2.0 SiO$_x$ compounds, respectively.

A coordination environment containing N atoms was represented as a fully connected graph $G = (X, A, S)$ in tuple form. Here, $X = (x_1, x_2, ..., x_N) \in \mathbb{R}^{N \times 3}$ denotes the cartesian coordinates of each atom with the excited oxygen site taken as the origin; $A \in \{0,1\}^{N \times 2}$ encodes the elemental species of each atom using a one-hot representation; and $S = (s, 0, ..., 0) \in \mathbb{R}^{N \times 200}$ holds the spectral information. In this notation, $s$ is the spectral vector at the excited oxygen site obtained from the calculations described above, and $0$ denotes a zero vector of the same size as $s$. The first indices of $X, A$, and $S$ are assigned to excited oxygen; there are no constraints on how the remaining indices are assigned.

*Model architecture*

An equivariant graph neural network (GNN) that incorporates the inductive bias of equivariance was used because it is suitable for molecular and crystalline structure generation[25,26,48,49]. In EGNN[48], the message-passing mechanism learns message vectors that encode interactions between neighboring nodes and aggregates them to update each node's coordinates $x$ and features $h$. In our implementation, we concatenate the spectral vector $s$ to the node features $h$, so that the learned messages—and consequently the updates of $x$ and $h$—are conditioned on the spectral information.

*Evaluation of coordination environments*

Next, we describe the evaluation methodology for the predicted coordination environments. The predicted coordinates were assessed using the RMSD. Because the model architecture is an equivariant graph neural network, the orientation of the generated graph and the ordering of node indices are not uniquely determined. Therefore, prior to coordinate evaluation, we perform alignment between the target structure and the generated structure. For alignment, we use the Kabsch algorithm[50,51], which computes the rotation matrix that minimizes the RMSD between two point sets when the corresponding point pairs are known. However, because the index order in the generated graph is not fixed, the correspondence between points is unknown. To manage computational cost, we handle alignment differently depending on whether the number of atoms is smaller than 9 or not.

For fewer than 9 atoms, we search all N! atoms; for structures with 9 or more atoms, we first consider a subset composed of the excited oxygen site and the seven nearest atoms (eight atoms in total). For this subset, we evaluate all 8! pairings using the Kabsch algorithm and rotate the original structure using the rotation matrix that gives the minimum RMSD. We then compute a cost matrix between the two full point sets and use the Hungarian algorithm to determine the pairing that minimizes the total cost. Using this pairing, we run the Kabsch algorithm again to obtain the final rotation matrix for alignment.

**Data availability**
The structural data used for constructing the dataset and for spectrum calculations were obtained from Materials Project (http://www.materialsproject.org)[42,43]. The dataset of calculated O-$K$ spectra is available upon request to corresponding authors.

**Code availability**
The code is available on request to corresponding authors.

**Acknowledgements**
We would like to acknowledge to Prof. Kiyou Shibata, Nagoya University for his helpful discussion. This study is supported by JSPS-KAKENHI, JST, and NEDO.

**Author Contributions**
R.O, I.T. and T.M. designed the research. R.O. performed the calculations and analyzed the data. Y.F. calculated the amorphous Si-O structures, and R.O., I.T., Y.F. and T.M. wrote the manuscript.

**Competing Interests**
The authors declare no competing interests.

# "Generative Inverse Estimation of 3D Atomic Coordination from Near-Edge Spectra via Equivariant Diffusion Models"

[1]Ren Okubo, [1,2]Yu Fujikata, [1]Izumi Takahara, and [1]Teruyasu Mizoguchi

[1]Institute of Industrial Science, The University of Tokyo,
Komaba, Meguro, Tokyo 153-8505, Japan

[2]Science & Innovation Center, Mitsubishi Chemical Corporation,
Kanagawa, 227-8502, Japan

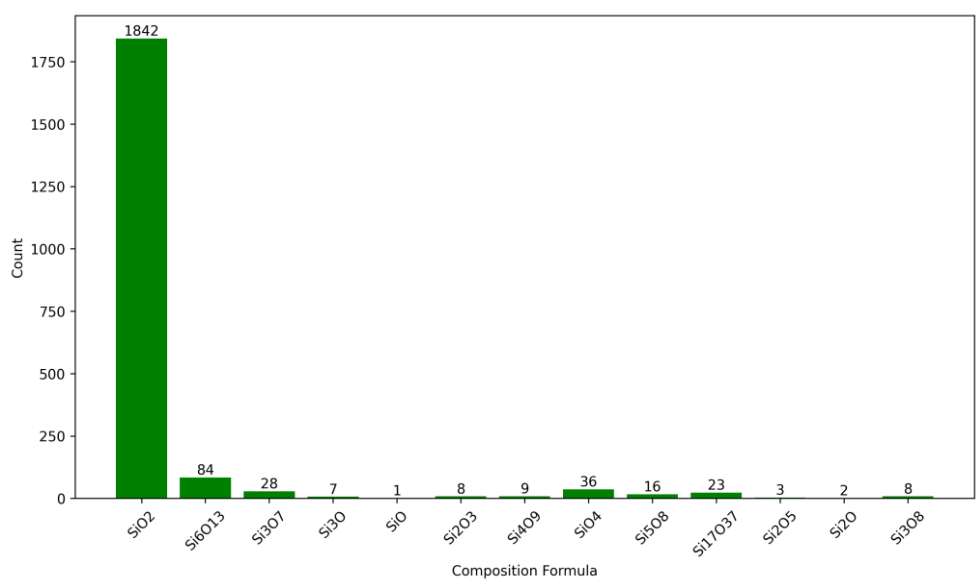

**Supplementary Figure 1:** Distribution of composition in crystalline 274 Si-O dataset obtained from Materials Project. Based on crystal structure, 2,067 oxygen K edge spectra were calculated from symmetrically unique oxygen sites from those Si-O dataset.

**Supplementary Figure 2.** Result of atom type prediction for $SiO_{0.25}$, $SiO_{2.0}$ amorphous

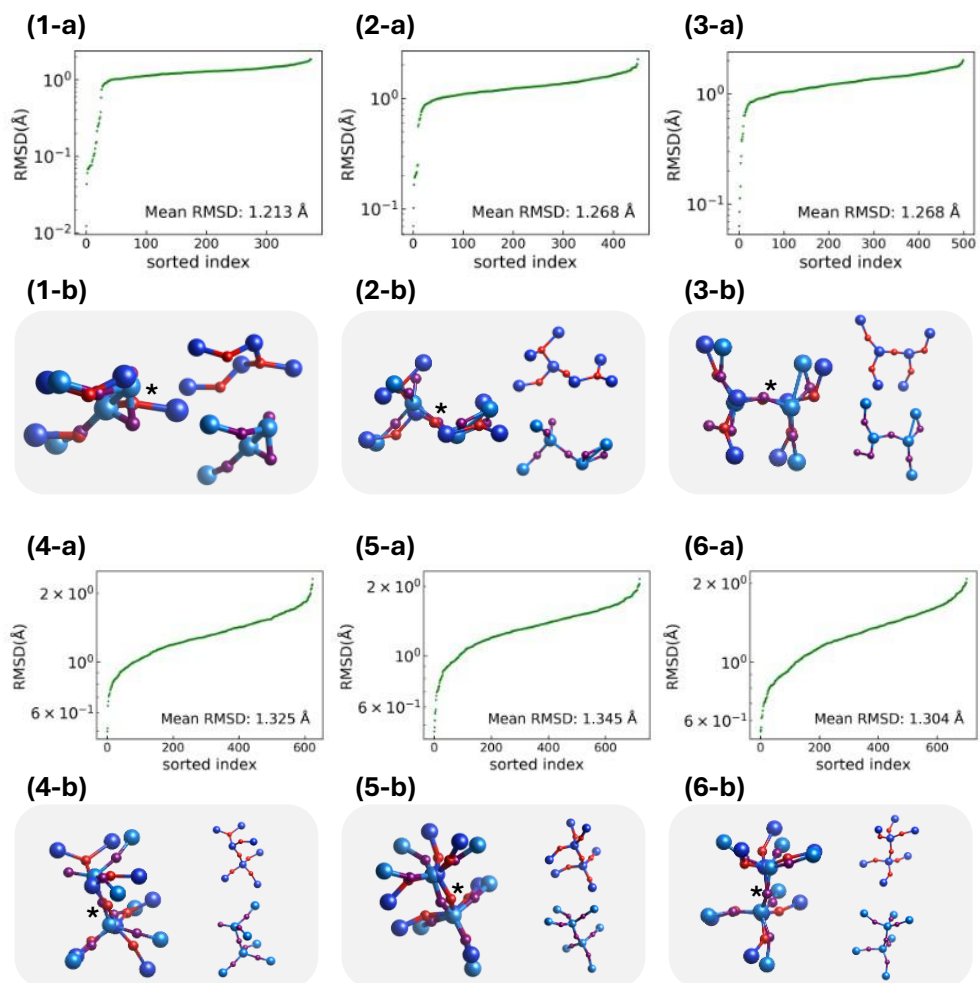

**Supplementary Figure 3 Conditional Generation for Amorphous.** (a)Coordination evaluation for each amorphous. (1-6) shows the result for $SiO_{0.5}$, $SiO_{0.75}$, $SiO_{1.0}$, $SiO_{1.25}$, $SiO_{1.5}$, $SiO_{1.75}$, respectively. (b) Target structures and predicted structures from spectrum of each amorphous. The target structure is shown in the upper right of each frame and the generated structure in the lower right; an overlay of these two structures is shown to the left. The asterisk-marked site corresponds to the excited atomic site. Blue and red spheres correspond to Si and oxygen, respectively.

**Supplementary Text 1 Hyperparameter settings**

We employed an equivariant graph neural network (EGNN) in this study. The EGNN performs message passing internally, and we set the number of message-passing layers $L$ to 5. For the neural network used in message passing, the hidden dimension was set to 1024, and the message vector dimension was set to 256.

For the forward diffusion process, following Hoogeboom et al. (2022), we defined the diffusion according to:

$$x_t = \alpha_t x_0 + \sigma_t \epsilon,$$
$$\alpha_t = (1 - 2s)\left(1 - \left(\frac{t}{T}\right)^2\right)^p + s,$$
$$\sigma_t = \sqrt{1 - \alpha_t^2}.$$

Here, we set $s = 1.0 \times 10^{-5}$ and $p = 2$. $T$ denotes the total number of timesteps in the diffusion process, and we used $T = 1000$.

Training was performed for 3000 epochs with a batch size of 1. We used AdamW as the optimizer with a learning rate of $1.0 \times 10^{-4}$ and a weight decay of $1.0 \times 10^{-12}$. During training, we updated the model parameters $\theta_t$, and at inference time we used the exponential moving average (EMA) parameters $\theta_t^{\text{EMA}}$. The EMA parameters were computed as:

$$\theta_t^{\text{EMA}} = \beta \theta_{t-1}^{\text{EMA}} + (1 - \beta)\theta_t,$$

with $\beta = 0.999$.